# EXCITED NEUTRINO AT NEXT LINEAR COLLIDERS

*NUCLEAR PHYSICS INSTITUTE, MOSCOW STATE UNIVERSITY*
*119899 MOSCOW, RUSSIA*

The possibility of single and pair excited neutrino production in high energy $e^+e^-$, $\gamma e$ and $\gamma\gamma$ collisions on linear colliders is studied. The integrated cross sections of these subprocesses are calculated. Special attention is paid to search for excited neutrino in $\gamma e^- \to W^- W^+ e^-$ process. Lower limits for the compositeness parameter estimated which will be available on the experiments at VLEPP, SLAC, JLC and DESY future linear colliders.

## 1. Introduction

Standard Model (SM) at present is in a good agreement with present experimental data. But from the theoretical point of view SM has a number of shortcomings and thus can not be considered as a complete theory of elementary particles.

The natural scale of the possible "new" physics wich comes, in particular, from the analysis of W and Z longitudinal components scattering amplitudes, from the analysis of the $g-2$ properties of electron and muon, and from the estimation of quark and lepton radii (high energy $e^+e^-$ and qq collisions) is a value of $\Lambda$ 1 TeV.

At present great attention is paid to study of possible compositeness of leptons and quarks. One of the signals of compositeness independent of concrete model should be the existing of the excited fermion states. Masses of the excited quark and lepton states are expected to be of order $\Lambda$ and can be in the interval $0.1 - 1$ TeV.

Analysis of experiments at present energies gives restrictions on the possible scale of the compositeness of the excited leptons shown in Table 1 (see [1] ):

Table 1: Scale limits for contact interactions.

| Type | Value(TeV) | Cl(%) | collab. |
|---|---|---|---|
| $\Lambda_{LL}^+$(eeee) | $> 1.4$ | 95 | 88 TASSO |
| $\Lambda_{LL}^-$(eeee) | $> 3.3$ | 95 | 88 TASSO |
| $\Lambda_{LL}^+$(ee$\mu\mu$) | $> 4.4$ | 95 | 86 JADE |
| $\Lambda_{LL}^-$(ee$\mu\mu$) | $> 2.1$ | 95 | 86 JADE |
| $\Lambda_{LL}^+$(ee$\tau\tau$) | $> 2.2$ | 95 | 86 JADE |
| $\Lambda_{LL}^-$(ee$\tau\tau$) | $> 3.2$ | 95 | 86 JADE |
| $\Lambda_{LR}^\pm$($\mu\nu_\mu e\nu_e$) | $> 3.10$ | 90 | 86 LBL,NWES TRIUMF |
| $\Lambda_{LL}^+$(eeqq) | $> 1.7$ | 95 | 91 CDF |
| $\Lambda_{LL}^-$(eeqq) | $> 2.2$ | 95 | 91 CDF |
| $\Lambda_{LL}^+$($\mu\mu$qq) | $> 1.4$ | 95 | 92 CDF |
| $\Lambda_{LL}^-$($\mu\mu$qq) | $> 1.6$ | 95 | 92 CDF |
| $\Lambda$(qqqq) | $> 0.825$ | 95 | 91 UA2 |

On the other hand analysis of experiments at LEP imposes restrictions on the masses of the excited charged leptons [1] given in Table 2.

There is a large number of papers which consider the excited lepton production in $e^+e^-$ collisions (for example, [2, 3]). But it is necessary to point out that colliders of alternative type – $\gamma e$ and $\gamma\gamma$ colliders – give new possibilities for investigation of some physical phenomena (production of excited leptons and quarks, color excitation of Z bosons, Higgs production, polarization phenomena) compared to $e^+e^-$ colliders. In particular, the most preferable reaction for search of the excited electron is $\gamma e \to e^*$ [3]. This type of colliders based on using of the high energy photons generated by Compton back-scattering of laser light [4]. Practical realization of $\gamma e$ and $\gamma\gamma$ colliders based on the corresponding $e^+e^-$ linear colliders is under consideration in the framework of VLEPP, JLC and SLAC projects.

Table 2: Mass limits for excited charged leptons. ($\lambda = m_l \Lambda / \Lambda$)

| Type | Value(GeV) | Cl(%) | Comment | $\lambda_Z$ |
|---|---|---|---|---|
| $e(e^*)$ | $> 91.0$ | 95 | $e^+e^- \to Z \to e^*e$, 92 ALEPH | $> 1$ |
|  | $> 44.6$ | 95 | $e^+e^- \to Z \to e^*e^*$, 92 ALEPH | $> 1$ |
|  | $> 116$ | 95 | from $e^+e^- \to e^* (t-\text{channel}) \to ee$, | $> 1$ |
|  |  |  | 91 OPAL(indirect effect) | $> 1$ |
| $\mu(\mu^*)$ | $> 91.0$ | 95 | $e^+e^- \to Z \to \mu^*\mu$, 92 ALEPH | $> 1$ |
|  | $> 46.1$ | 95 | $e^+e^- \to Z \to \mu^*\mu^*$, 92 ALEPH | $> 1$ |
| $\tau(\tau^*)$ | $> 90.0$ | 95 | $e^+e^- \to Z \to \tau^*\tau$, 92 ALEPH | $> 1$ |
|  | $> 46.0$ | 95 | $e^+e^- \to Z \to \tau^*\tau^*$, 92 ALEPH | $> 0.18$ |

Some theoretical papers consider excited lepton production at $e^+e^-$, pp [3], $\gamma e$ and $\gamma\gamma$ [5] future colliders. The main results of [3, 5] are the following: if compositeness of leptons is realized at a scale $\leq O(10\text{ TeV})$, excited lepton states can be found in $e^+e^-$ and $\gamma e$ collisions with masses up to maximal available energy $\sqrt{s}$. In [3] it is also said that the lower limit for excited charged leptons at LHC can be reached up to 4 TeV ($\sqrt{s} = 16$ TeV, $\Lambda \leq 10$ TeV). Interesting aspects of excited neutrino production at Z pole ($Z \to \nu_e \nu_e^*$, LEP I) was a subject of paper [6]. This paper gives the following upper limit on the possible masses of the excited neutrino: up to 90 GeV if $\lambda = m_l \Lambda / \Lambda = 0.09$.

Here we consider the least studied (compared to charged lepton production) excited neutrino production in more detail in $e^+e^-$, $\gamma e$ and $\gamma\gamma$ collisions in the TeV energy region. These excited neutral states are supposed to be the isospin 1/2 partners to the much studied excited charged leptons and have spin 1/2 in the most simple realization of models (higher spin and isospin assignments have been also discussed elsewhere [7]). Transition between the ordinary light fermions and the excited ones and also between both excited fermions are described in SU(2)⊗U(1) invariant form by the following effective Lagrangian [8]:

$$L^1_{\text{eff}} = \bar{\ell}^* \gamma^\mu (g \frac{\emptyset}{2} W_\mu + g' \frac{Y}{2} B_\mu) \ell^*, \quad (1)$$

$$L^2_{\text{eff}} = \frac{1}{2\Lambda} \bar{\ell}^*_R \sigma^{\mu\nu} (fg \frac{\emptyset}{2} W_{\mu\nu} + f' g' \frac{Y}{2} B_{\mu\nu}) \ell_L + \text{h.c.}, \quad (2)$$

where $g$ and $g'$ are the usual SU(2) and SU(1) weak couplings; $\emptyset$ denotes the Pauli matrices; $\ell^*$ and $\ell$ are isodoublets of the excited and usual leptons respectively; constants $f$ and $f'$ are supposed to be of order unity.

Using this model we consider the following processes of single and pair production of the excited neutrino:

$$e^+ e^- \to \nu^* \bar{\nu} \quad (3)$$

$$\gamma e^- \to \nu^* W^- \quad (4)$$

$$e^+ e^- \to \nu^* \bar{\nu}^* \quad (5)$$

$$\gamma\gamma \to \nu^* \bar{\nu}^* \quad (6)$$

For processes (3), (4) and (5) we put $f = f' = 1$, while for process (6) we put $f = 1$, $f' = -1$, because the $(f - f')$ factor for the vertex $\gamma \nu_e \bar{\nu}_e^*$ eliminates it in the case of $f = f' = 1$. Moreover, the analysis of the future experiments on $\gamma\gamma$ collisions would be very important for the determination of $f$ and $f'$, because the cross section in this case is proportional to $(f - f')^2$.

The main question which is discussed here is the sensitivity of the future experiments in the TeV region to the signals from the excited neutrino certainly it is connected with the limits on the values of $\Lambda$ and $m_{\nu_e}\Lambda$.

It should be pointed out that all calculations here both analytical and numerical have been done with the aid of CompHEP software package [9], the present version of which performs the following principal operations:
1. analytic calculations of squared matrix elements (with the aid of Feynman rules) of the process $1 \to 2$, $2 \to 2$, $2 \to 3$ in the first order of any input model;
2. numeric integration of such processes and graphic presentation of the final results (for $2 \to 3$ processes with the help of BASES integrated pakage wich is connected with CompHEP through special interface).

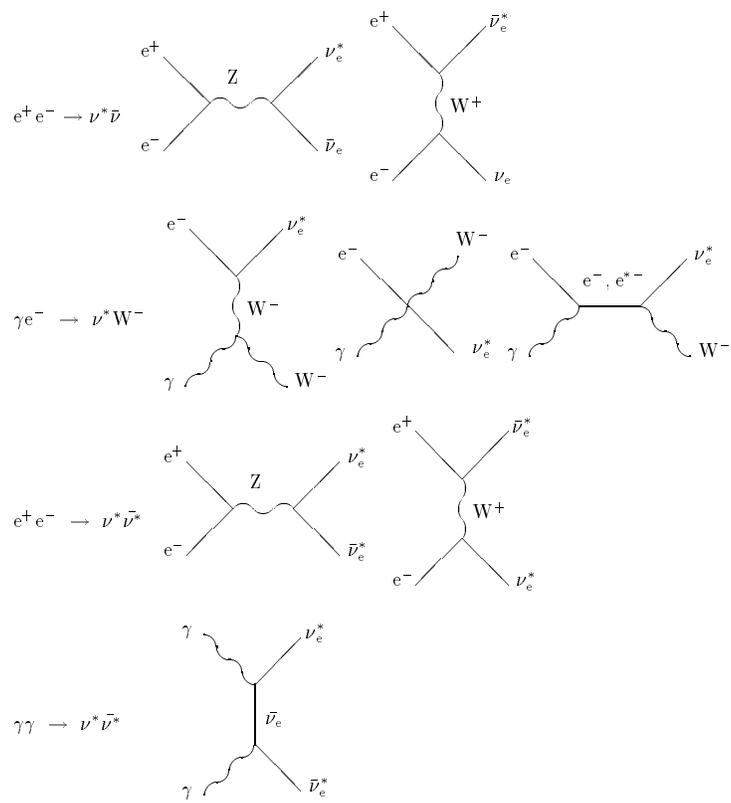

## 2. Decay modes

Decay widths and thus the branching ratios for excited neutrino which could be obtained by straightforward calculations from (2) and represented by the following expression :

$$\Gamma(\nu_e^* \to \nu_e(e)V) = \frac{\alpha}{4} \frac{m_{\nu_e\Lambda}^3}{\Lambda^2} f_V^2 \left(1 - \frac{M_V^2}{m_{\nu_e\Lambda}^2}\right)^2 \left(1 + \frac{M_V^2}{2m_{\nu_e\Lambda}^2}\right), \tag{7}$$

where

$$V = \gamma, Z \text{ or } W; \quad f_\gamma = \frac{f - f'}{2}, \quad f_Z = \frac{f c_W^2 + f' s_W^2}{2 s_W c_W}, \quad f_W = \frac{f}{\sqrt{2} s_W},$$

$s_W$ and $c_W$ being sin and cos of Weinberg angles. In Tables 3 we present decay modes and the corresponding branching ratios of the excited neutrino $\nu_e^*$ (for $m_{\nu_e\Lambda}$=0.1, 0.5 and 1.0 TeV) when $f = 1$, $f' = 1$ and $f = 1$, $f' = -1$.

Table 3: Branching ratios for the excited neutrino decay ($f = 1$).

| $m_{\nu_e\Lambda}$ (GeV) | Branching ratios | | | | | | Total decay width(GeV) | |
|---|---|---|---|---|---|---|---|---|
| | $\nu_e^* \to eW$ | | $\nu_e^* \to \nu Z$ | | $\nu_e^* \to \nu\gamma$ | | | |
| | $f = f'$ | $f = -f'$ | $f = f'$ | $f = -f'$ | $f = f'$ | $f = -f'$ | $f = f'$ | $f = -f'$ |
| 100 | 0.866 | 0.268 | 0.134 | 0.012 | 0 | 0.720 | 0.00084 | 0.0027 |
| 500 | 0.610 | 0.602 | 0.390 | 0.115 | 0 | 0.283 | 0.85 | 0.86 |
| 1000 | 0.608 | 0.606 | 0.392 | 0.117 | 0 | 0.277 | 7.0 | 7.1 |

It is seen from Table 3 that dominating decay channel is $\nu_e^* \to eW$. So the more preferable method of searching of the excited neutrino through its decay into electron and W-boson, which decay then in two jets (it is dominating decay channel of W-boson with branching 80 %).

## 3. Pair and single excited neutrino production

### 1. General remarks

It is worth noting that for $e^+e^-$ colliders with c.m. energies 0.5, 1.0 and 2.0 TeV the respective effective energies for $e\gamma$ and $\gamma\gamma$ collisions will be approximately the following: 0.375, 0.750, 1.5 TeV for $e\gamma$ collisions (i.e. $\sqrt{s}_{e\gamma} = 0.75\sqrt{s}_{e^+e}\Gamma$) and 0.350, 0.700, 1.4 TeV for $\gamma\gamma$ collisions (i.e. $\sqrt{s}_{\gamma\gamma} = 0.7\sqrt{s}_{e^+e}\Gamma$). Of course strictly speaking it is necessary to fold the cross section with the real $\gamma$-beam energy spectrum and it will be done in future, but in the main approximation it is convenient to use such rescaling of $\sqrt{s}$.

The dependence of the calculated integrated cross section $\sigma$ on $m_{\nu_e\Lambda}$ or $\Lambda$ when one of the alternative variables is fixed is shown in Figs. 1, 2 ($\sqrt{s}_{e\gamma} = 0.75\sqrt{s}_{e^+e}\Gamma$ and $\sqrt{s}_{\gamma\gamma} = 0.7\sqrt{s}_{e^+e}\Gamma$).

From the point of view of building a realistic model more preferable case is $m_{\nu_e\Lambda} \sim \Lambda$. The case when $m_{\nu_e\Lambda} \ll \Lambda$ requires considering some special mechanisms to explain such difference of $m_{\nu_e\Lambda}$ from $\Lambda$ and thus to make this model natural. It is assumed that $\Lambda$ is of order of one TeV. But we should not disregard the case when excited lepton mass is of order of one hundred GeV, because we know the mechanism by which the possible lepton constituents could be bound.

## 2. Single excited neutrino production

Let us consider the processes of single production of the excited neutrino — $e^+e^- \to \nu^*\bar{\nu}$ and $\gamma e^- \to \nu^* W^-$ (5) processes and compare the behavior of $\sigma$ which dependes on $\Lambda$ and $m_{\nu_e}\Lambda$ in these two cases. The corresponding total cross sections are :

$$\sigma(e^+e^- \to \nu^*\bar{\nu}) = \frac{\pi\alpha^2 f^2}{4\Lambda^2}\beta\bigg\{\frac{1}{s_W^4}\bigg[(2w+\beta)\log(1+\frac{\beta}{w}) - 2\beta\bigg]$$

$$+ \frac{1-2s_W^2}{(1-z)s_W^4 c_W^2}\bigg[w(1+\frac{w}{\beta})\log(1+\frac{w}{\beta}) - \frac{1}{2}(\beta+2w)\bigg] + \frac{1+(1+4s_W^2)^2}{16(1-z)^2 s_W^4 c_W^4}\beta(1-\frac{2}{3}\beta)\bigg\}, \quad (8)$$

$$\sigma(\gamma e^- \to \nu^* W^-) =$$

$$\frac{\pi\alpha^2 f^2}{8s_W^2 \Lambda^2}\bigg[4\log\frac{a-R-w-1}{a+R-w-1}\bigg(-2a^3 + (3w+2)a^2 - (w+4)a - w^3 - w^2 + \frac{4w+2}{1-a}\bigg)$$

$$+ R\bigg(-14a^2 + (7w-2)a + (7w^2+13w-26) + \frac{(14w-16)a - 18w+8}{a^2-2a+1}\bigg)\bigg], \quad (9)$$

where $a = m_{\nu_e}^2\Lambda/s$ $(m_{\nu_e}\Lambda = m_e\Lambda)$, $\beta = 1 - m_{\nu_e}^2\Lambda/s$,

$$w = M_w^2/s, \quad z = M_z^2/s, \quad R = \sqrt{1+a^2+w^2-2aw-2a-2w}.$$

Dependence of the integrated cross section $\sigma$ on $\Lambda$ obviously looks like $(1/\Lambda)^2$ (Figs. 1a, 1b). The dependence of $\sigma$ on the excited neutrino mass $m_{\nu_e}\Lambda$ is shown on Figs. 2a, 2b. Of course, the maximum available $m_{\nu_e}\Lambda$ is higher in case (3), because of the effective c.m. energy would be slightly higher in the case of $e^+e^-$ collisions ($\sqrt{s}_{e\gamma} = 0.75\sqrt{s}_{e^+e}\Gamma$). But if we consider $e^+e^- \to \nu^*\bar{\nu}$ and $e^+e^- \to \nu^*\bar{\nu}^*$ and compare values of $\sigma$ in the range of $m_{\nu_e}\Lambda$ not very close to the kinematical limit $\sqrt{s}_{\gamma e}$, then for (5) $\sigma$ is about one order higher then for (3). This is due to the fact that in (5) the vertex factor for $\gamma WW$ is proportional to the momenta of the particles in the vertex and thus gives a higher contribution to the integral compared to that of the $eW\nu$. For instance, $\sigma = 0.6$ pb for (5) and $\sigma = 0.08$ pb for (3), when $\Lambda = 3$ TeV and $m_{\nu_e}\Lambda = 0.5$ TeV at $\sqrt{s}_{e^+e}\Gamma = 1.5$ TeV ($m_e\Lambda = 2$ TeV in the s-channel). In terms of the number of events we shall have 3600 for (5) and 480 for (3) events of the excited neutrino decay taking into account the dominant decay branching $\mathrm{Br}(\nu_e^* \to eW) \simeq 0.6$ at the luminosity $10^4 \mathrm{pb}^{-1}$. Therefore the maximum detectable value of $\Lambda$ is higher in the case (5).

One can see that in case (5) $\sigma$ depends much stronger on $\sqrt{s}$ and on $\Lambda$ compared to case (3).

It is necessary to point out that all analytical expressions considered above coincide with the formulas in the independent parallel paper [10].

## 3. Excited neutrino pair production

Now let us turn to the pair production of the excited neutrino. Analytical expressions for total cross section for these processes are given by

$$\sigma(e^+e^- \to \nu^*\bar{\nu}^*) = \frac{\pi\alpha^2 s}{96 s_W^4 c_W^4 (1-z)^2}\beta(3-\beta^2)((4s_W^2-1)^2+1), \quad (10)$$

$$\sigma(\gamma\gamma \to \nu^*\bar{\nu}^*) = \frac{\pi\alpha^2 f^2 s}{4\Lambda^4}\beta^2\bigg[(1-\beta^2)^2 \log\frac{1+\beta}{1-\beta} + 2\beta(\frac{5}{3}-\beta^2)\bigg](f=-f'), \quad (11)$$

where $\beta = \sqrt{1-m_{\nu_e}^2\Lambda/s}$.

For process $e^+e^- \to \nu^*\bar{\nu}^*$ we present the cross section only for s-channel Z-boson exchange (which gives the main contribution) in order not to overload this paper with comparatively large expressions.

It is seen from Fig. 1c that $\sigma$ for $e^+e^- \to \nu^*\bar{\nu}^*$ becomes independent of $\Lambda$ at $\Lambda > \sqrt{s}$. while for $\gamma\gamma \to \nu^*\bar{\nu}^*$ it behaves like $(1/\Lambda)^4$ (Fig. 1d). This happens because at large $\Lambda$ the main contribution to the cross section for (5) is given by the vertex with two excited neutrinos: $Z\nu^*\bar{\nu}^*$ (which is independent on $\Lambda$). Dependence of the cross section on the excited neutrino mass for both processes is shown in Figs. 2c, 2d. One can see that $\sigma$ is much higher for $e^+e^- \to \nu^*\bar{\nu}^*$ compared to $\gamma\gamma \to \nu^*\bar{\nu}^*$. Hence the sensitivity of the experiment on excited neutrino pair production will be higher in the case of $e^+e^-$ collisions and thus the achievable values of $m_{\nu_e}\Lambda$ and $\Lambda$ in the case (4) are higher.

## 4. Numerical results

Our numerical results which show measurable values of $\Lambda$ for different energies of planned colliders are summarized in Table 4 (here we assume the integrated luminosity per year as high as $10^4$ pb$^{-1}$ and 100 events excited neutrino production per year with decay channel $\nu_e^* \to eW$ as a criterion for observability of the effect). Note that $\sigma$ for $e^+e^- \to \nu^*\bar{\nu}^*$ becomes independent on $\Lambda$ and so we do not present corresponding values.

Table 4: The values of $\Lambda$ which are measurable when $\sqrt{s}_{e^+e^-} = 0.5, 1$ and 2 TeV.

| | $e^+e^- \to \nu^*\bar{\nu}$ | | $\gamma e^- \to \nu^*W^-$ | | $e^+e^- \to \nu^*\bar{\nu}^*$ | | $\gamma\gamma \to \nu^*\bar{\nu}^*$ | |
|---|---|---|---|---|---|---|---|---|
| $\sqrt{s}_{e^+e^-}$ (Gev) | $m_{\nu_e}$ (GeV) | $\Lambda$ (TeV) | $m_{\nu_e}$ (GeV) | $\Lambda$ (TeV) | $m_{\nu_e}$ (GeV) | $\Lambda$ (TeV) | $m_{\nu_e}$ (GeV) | $\Lambda$ (TeV) |
| 500 | 100 | 6.4 | 100 | 9.1 | < 175 | I | 100 | 0.7 |
| | 200 | 5.5 | 200 | 7.4 | | | 200 | X |
| | 400 | 2.0 | 400 | X | | | 400 | X |
| 1000 | 100 | 8.5 | 100 | 22 | < 350 | I | 100 | 1.0 |
| | 200 | 8.2 | 200 | 21 | | | 200 | 0.95 |
| | 400 | 7.0 | 400 | 19 | | | 400 | X |
| 2000 | 100 | 10.5 | 100 | 92 | < 350 | I | 100 | 1.4 |
| | 200 | 10.2 | 200 | 91 | | | 200 | 1.3 |
| | 400 | 9.7 | 400 | 86 | | | 400 | 1.2 |

In this table X means that the corresponding particle can not be produced while I means the cross section independent on $\Lambda$ and thus the respective values of compositeness parameter are'not presented. The tables and analysis considered above clearly show that $\gamma e^- \to \nu^*W^-$ process is more preferable for single excited neutrino production studies in comparison with $e^+e^- \to \nu^*\bar{\nu}$ while for the pair production $e^+e^- \to \nu^*\bar{\nu}^*$ is more profit then $\gamma\gamma \to \nu^*\bar{\nu}^*$.

It is obvious that our estimations are quite rough and for more precise analysis of (3)–(6) processes it is necessary to consider the main modes of decay of the exited neutrino in $2 \to 3$ processes, to fold the subprocesses cross sections with the real photon spectrum and to compare total cross sections with the standard model predictions. All this for the $\gamma e^- \to W^-\nu^* \hookrightarrow W^+e^-$ process we try to do in the next section of this paper.

### 4. $\gamma e^- \to W^-W^+e^-$ process: searching for signal from excited neutrino

In this section we present more realistic analysis of excited neutrino production. The main contribution to the deviation of integrated cross section for considered process from Standard Model comes from diagrams $\gamma e^- \to W^-\nu^* \hookrightarrow W^+e^-$. So, it is necessary to consider invariant mass distribution $d\sigma/dM$, where M - invariant mass of outgoing $e^-$ and $W^+$. Results of the numerical calculations are presented in Fig. 3. These calculation performed at $\Lambda = 3$ TeV, $\sqrt{s} = 0.5$, 1, 1.5 and 2 TeV. There are two pair distributions in the Fig. 3. First kind represented calculation without folding $\sigma$ with the photon spectra: dot-dashed line shows invariant mass distribution according to the Standard Model while dotted histogram show distribution when excited neutrino is produced. Dashed line and solid line histogram represent respective distributions which are folded with photon spectra ($m_{\nu_e}\Lambda$ = 0.3, 0.6, 1 and 1.4 TeV for $\sqrt{s}$= 0.5 (Fig. 3a), 1 (Fig. 3b), 1.5 (Fig. 3c) and 2 (Fig. 3d) TeV respectively). The widths of histogram bins are equal to 10 GeV for $\sqrt{s}$ = 0.5 and 1 TeV; 20 GeV for $\sqrt{s}$ =1.5 TeV and 50 GeV for $\sqrt{s}$ =2 TeV. The bins outside resonance region are merged. The choice of such bin width seems reasonable because the energy resolution (for example for JLC) is planned as following:

$$\frac{\sigma_E}{\sqrt{E}} = \frac{15\%}{\sqrt{E}} + 1\% \ . \tag{12}$$

Table 5: The main characteristic values for $\gamma\, e^- \to W^- W^+ e^-$ process and presented at Figure 3 histograms

| c.m. energy (TeV) | 0.5 | 1.0 | 1.5 | 2.0 |
|---|---|---|---|---|
| $m_{\nu_e}\Lambda$ (TeV) | 0.3 | 0.6 | 1.0 | 1.4 |
| energy resolution (GeV) | $\simeq 5$ | $\simeq 9$ | $\simeq 15$ | $\simeq 20$ |
| bin width of histogram (GeV) | 10 | 10 | 20 | 50 |
| number of events in resonance bin for processes: without $\nu_e^*$ | 200 | 170 | 220 | 400 |
| with $\nu_e^*$ | 240 | 320 | 500 | 750 |
| $\Gamma(\nu_e^* \to eW)$ $(f=f')$ (GeV) | 0.03 | 0.3 | 0.8 | 2.0 |

In Table 5 we represented the main characteristic values of the considered process and presented histograms. Shown in Fig. 3 is the case when resonance hits the middle of the bin. The signal exceeds 3 standard deviations from Standard model background (we assume integrated luminosity $10^4$ pb$^{-1}$ per year). In the worst case when the mass of resonance coincide with the border of bins the excess would be two standard deviations. But it is expected that real resolution will be of order of 1 GeV. So , in any case we would have a clear signal from excited neutrino.

In our calculations we made cut of the scattering angle of the electron: $10^o < \theta < 170^o$. It is worth mentioning that signal from the excited neutrino would be more clear if one optimize the angular cuts.

We have presented here results only fore $\Lambda = 3$ TeV just to demonstrate observability of the signal from excited neutrino. To learn the upper limits on $\Lambda$ which would be practically achievable at definite collider it is necessary to take into account specific properties of the specific experiment.

We are grateful to G. Jikia, Yu. Pirogov, P. Zerwas and A. Djouadi for fruitful discussions. We also wish to thank authors of the CompHEP system for valuable advice.

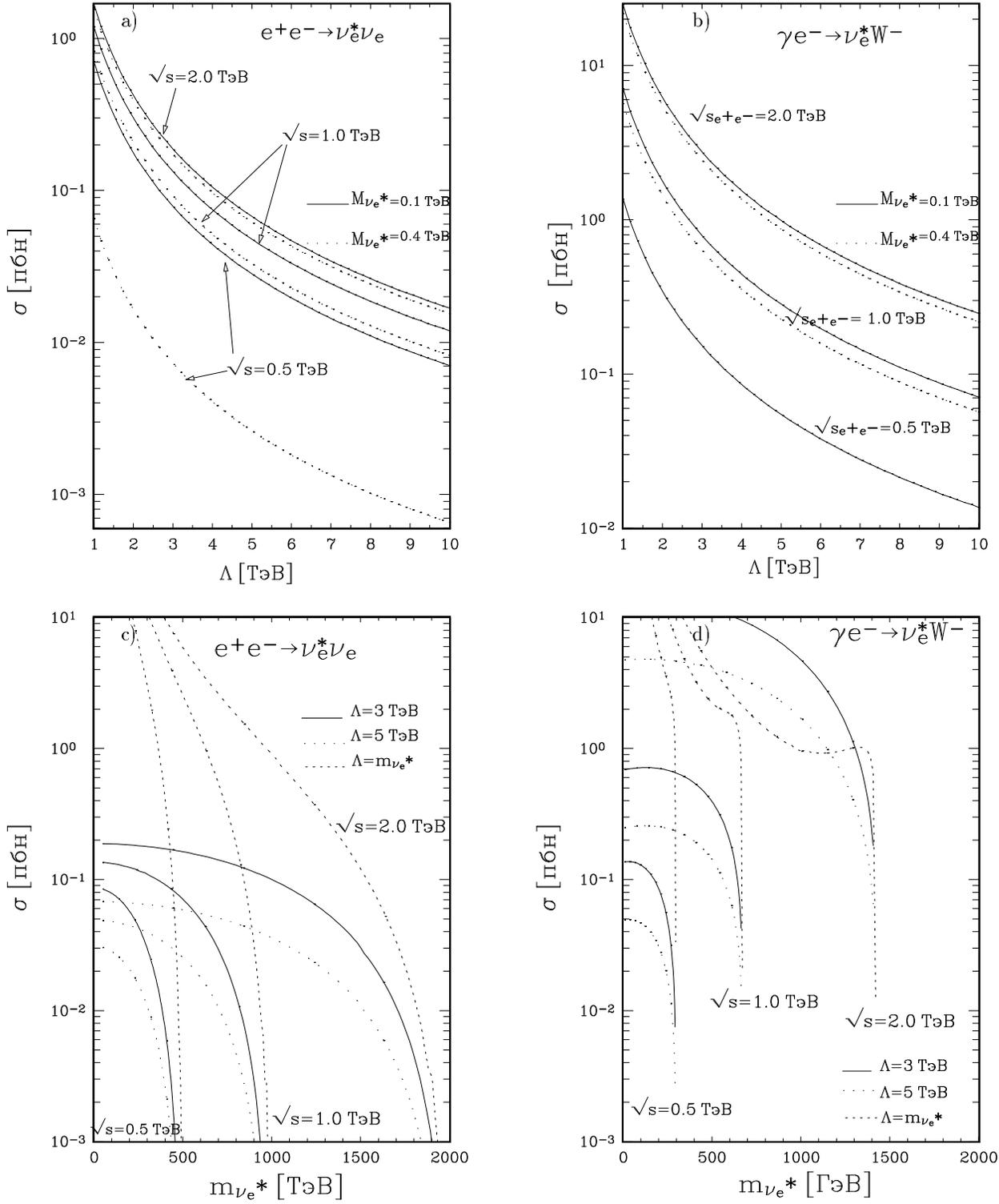

Figure 1: Total cross section versus $\Lambda$ ($\sqrt{s}_{e^+e^-} = 0.5,\ 1.0,\ 2.0$ TeV) for the processes: $e^+e^- \to \nu^*\bar{\nu}$(a), $\gamma e^- \to \nu^* W^-$(b) ($m_e \Lambda = 2$ TeV in the s-channel), $e^+e^- \to \nu^*\bar{\nu}^*$(c), $\gamma\gamma \to \nu^*\bar{\nu}^*$(d)

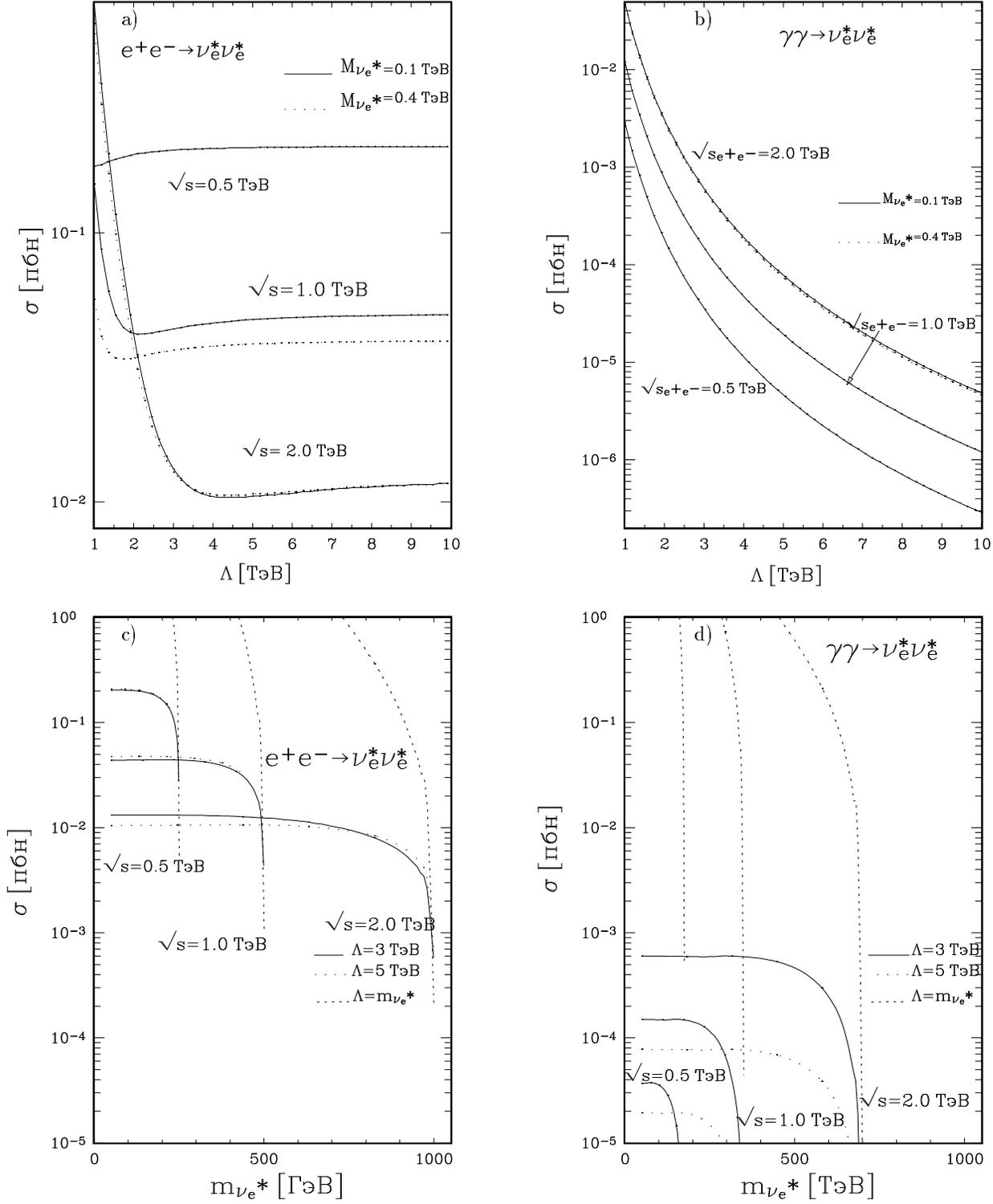

Figure 2: Total cross section versus parameter of compositeness $\Lambda$ ($\sqrt{s}_{e^+e^-} = 0.5, 1.0, 2.0$ TeV) for the processes: $e^+e^- \to \nu^*\bar{\nu}$(a), $\gamma e^- \to \nu^*W^-$(b) ($m_e\Lambda=2$ TeV in the s-channel), $e^+e^- \to \nu^*\bar{\nu}^*$(c), $\gamma\gamma \to \nu^*\bar{\nu}^*$(d)

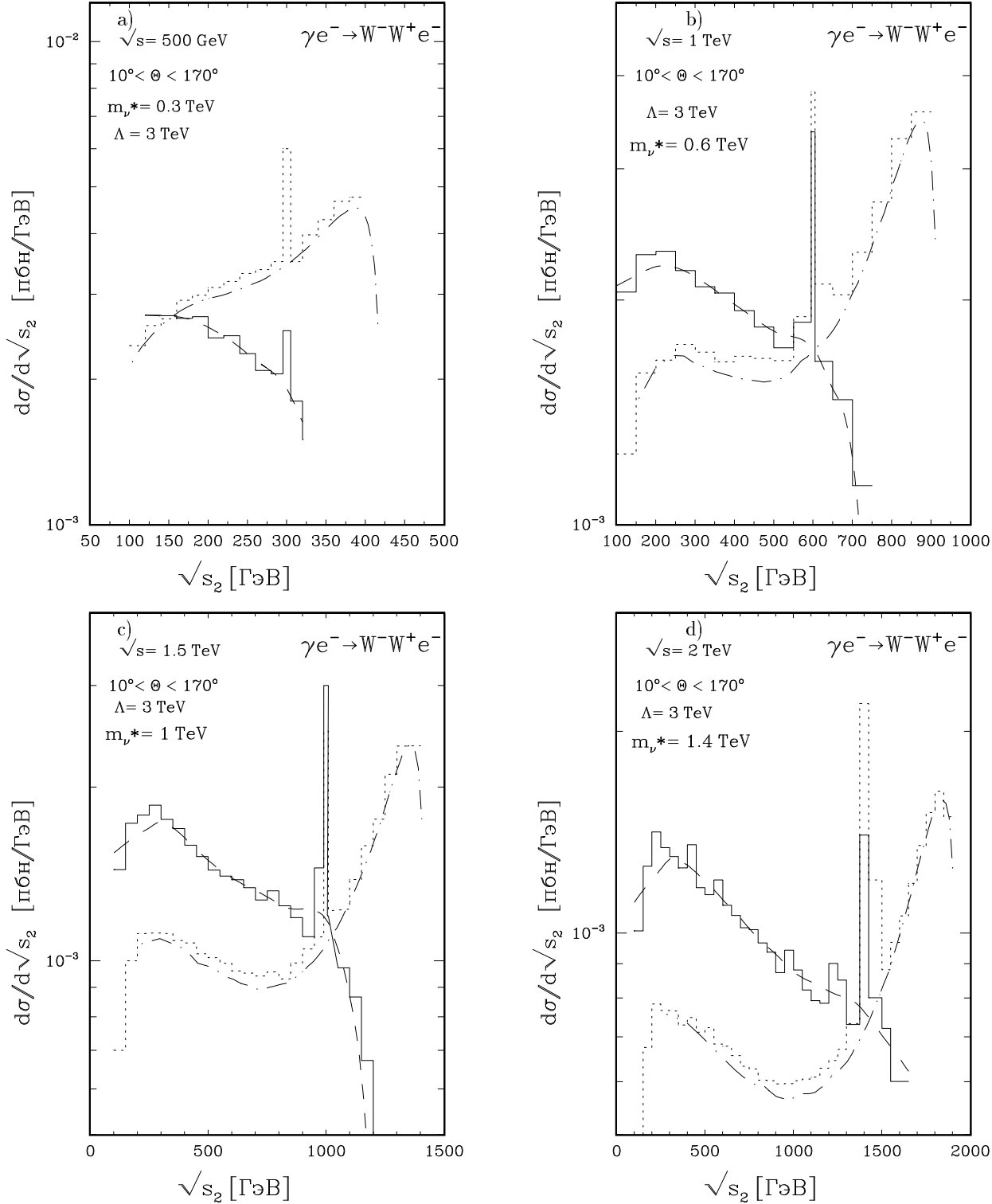

Figure 3: Invariant mass of outgoing $e^-$ and $W^+$ distribution for $\gamma e^- \to W^- W^+ e^-$ process. Calculation without folding with the photon spectra: dot-dashed line — invariant mass distribution according to the Standard Model; dotted histogram — distribution when excited neutrino is produced. Dashed line and solid line histogram represent respective distributions which are folded with photon spectra ($m_{\nu_e} \Lambda =$ 0.3, 0.6, 1 and 1.4 TeV for $\sqrt{s}=$ 0.5 (Fig. 3a), 1 (Fig. 3b), 1.5 (Fig. 3c) and 2 (Fig. 3d) TeV respectively)

11